\documentclass[
    a4paper,
    bibnotes,
    twocolumn,
    superscriptaddress,
    groupedaddress,
    floatfix,
    showpacs,
]{revtex4}
\usepackage{graphicx}
\usepackage{amssymb,amsmath}
\usepackage{amsmath}
\usepackage{color}

\begin{document}

\title{Current-induced forces and hot-spots in biased nano-junctions}
 \author{Jing-Tao \surname{L\"u}}  
\email{jtlu@hust.edu.cn}
\affiliation{School of Physics and Wuhan National High Magnetic Field Center, Huazhong University of Science and Technology, Wuhan, China}
  \affiliation{Department of Micro- and Nanotechnology, Technical
  University of Denmark, Kongens Lyngby, Denmark}
\affiliation{Niels Bohr Institute, Nano-Science Center, University of
  Copenhagen, Copenhagen, Denmark}
\author{Rasmus B. \surname{Christensen}}
  \affiliation{Department of Micro- and Nanotechnology, Technical
  University of Denmark, Kongens Lyngby, Denmark}
\author{Jian-Sheng \surname{Wang}}
\affiliation{Department of Physics, National University of Singapore, 117551 Singapore, Republic of Singapore}
\author{Per \surname{Hedeg{\aa}rd}} 
\affiliation{Niels Bohr Institute, Nano-Science Center, University of
  Copenhagen, Copenhagen, Denmark}

\author{Mads \surname{Brandbyge}}
  \affiliation{Center for Nanostructured Graphene (CNG), Department of Micro- and Nanotechnology, Technical
  University of Denmark, Kongens Lyngby, Denmark}

\begin{abstract}
We investigate theoretically the interplay of current-induced forces (CIF), Joule
heating, and heat transport inside a current-carrying nano-conductor. We find that the CIF, due to the electron-phonon coherence, can control the
spatial heat dissipation in the conductor.  This yields a significant
asymmetric concentration of excess heating (hot-spot) even for a symmetric
conductor. When coupled to the electrode phonons, CIF drive
different phonon heat flux into the two electrodes.  First-principles
calculations on realistic biased nano-junctions illustrate the importance of
the effect. 
\end{abstract}
	
\maketitle
\emph{Introduction--}
Current-induced forces and Joule heating both originate from the coupling between electrons and phonons\cite{note}, one of the most fundamental many-body
interactions responsible for a wide range of phenomena in
molecular and condensed-matter physics. Their vital role
in maintaining the electronic device stability is further
promoted at nanoscale. Our understanding of the two
closely related effects, especially their interplay in nano-
and atomic-conductors is still under development\cite{DuMcTo.2009,JMP.2010,BoKuEgVo.2011,ToDuDu.2011,Ps2011,Bustos2013,Schirm2013,Wheeler2013,Tchavdar14}.
Several forces, present only in the nonequilibrium situation, have been
discovered theoretically. Among them are the non-conservative (NC) ``wind
force'', and the Berry-phase (BP) induced pseudo-magnetic force.  Different
from the stochastic Joule
heating\cite{Ta.2006,GaRaNi.2007,GaRaNi.2008,HuChDa.2007,TsTaKa.2008,As08,SmNoUn.2002,WaLeKr.2004,KuLaPa.2004,IoShOp.2008,WaCoToNa.2010,Kaasbjerg13,Lee2013},
the NC and BP forces can generate \emph{deterministic} energy and momentum
transfer between the current-carrying electrons and the vibrations in the conductor\cite{DuMcTo.2009,JMP.2010,BoKuEgVo.2011,ToDuDu.2011,Ps2011}.  In
carefully designed devices, this effect may be used to drive atomic
motors\cite{DuMcTo.2009,Bustos2013}. Meanwhile, it can also impact the stability of the
device\cite{SmUnRu.2004,ScFrGa.2008,JMP.2010}.  To this end, the vibrational/phononic\cite{note} heat transport and heat distribution in the presence of current flow becomes an urgent problem to investigate.

The electrode phonons play an important role as heat sinks for the locally
dissipated Joule heat in the conductor\cite{TsTaKa.2008}.  However, the effects
on the heat transport of the deterministic CIF, and the
momentum transfer from the current has so far not been explored.  To address
this question, we go beyond the previous treatments\cite{JMP.2010,luprb12}
considering localized vibrations in the conductor, and include coupling to the
phonons in the electrodes\cite{wang07}. Employing the semi-classical generalized Langevin
equation(SGLE),\cite{FEYNMAN1963,CALE.1983,SC.1982,luprb12}, we find that, in
addition to energy transfer, the CIF also influence how the
excess vibrational energy is distributed in the junction and transported to the
electrodes. Using first-principles calculations, we demonstrate how
\emph{symmetric} current-carrying nano-junctions typically possess a
significant \emph{asymmetric} excess heat distribution with heat accumulation
at hot-spots in the junction.  At the same time the phonon heat flow to the two
electrodes differs. This behavior is governed by the phases of the electron and
phonon wavefunctions, and is a result of electron-hole pair symmetry breaking in
the electronic structure.  It will have important implications, and should be taken
into account when considering junction disruption at high bias\cite{SmUnRu.2004,Oshima2013}.

\emph{Method  --  }
In the SGLE approach we adopt the two-probe transport setup, where a ``bottleneck'' nano-junction (system) is connected
to left($L$) and right($R$) electrodes. We consider the case where the system region is characterized by a significant current density and deviation from equilibrium. The current-carrying electrons are treated as
a nonequilibrium bath, coupling linearly with the system displacement, while  
the remaining atoms in $L$ and $R$ form two phonon baths interacting with
the system also via a linear coupling. 
The electron-phonon (e-ph) coupling Hamiltonian can be written as
\begin{equation}
	\label{eq:eph}
	H_{eph} = \sum_{i,j,k}M_{ij}^k (c_i^\dagger c_j + h.c.)\hat{u}_k.
\end{equation}
Here, $\hat{u}_k = \sqrt{m_k} \hat{x}_k $ is the mass-normalized displacement away from the equilibrium
position of the $k$-th atomic degrees of freedom, with $m_k$ the mass, and $\hat{x}_k$ the displacement operator from equilibrium position; 
$c_i^\dagger$($c_j$) is the electron creation(annihilation) operator for the $i$-($j$-)th electronic
state in the junction. The coupling matrix, $M_{ij}^k$, is
local in real space, non-zero in the system and neglected in $L,R$. 
We treat the e-ph interaction perturbatively using the electron and phonon states obtained from the
Born-Oppenheimer approximation. In order to
focus on the effect of CIF, we will ignore the change of
Hamiltonian due to the applied voltage. 

The SGLE describing the dynamics of the \emph{system atoms} reads,
\begin{eqnarray}
	\ddot U(t)-F(U(t)) &=& -\int^t {\Pi^r}(t-t')U(t')dt' + f(t),
	\label{eq:langa}
\end{eqnarray}
where, $U$ is a vector composed of the mass-normalized displacements of the
system, and $F(U(t))$ is the force vector from the potential of the isolated system.
We adopt the harmonic approximation, $F(U(t))=-K U(t)$, with $K$ being the dynamical
matrix.  The effect of all bath degrees of freedom is hidden in the terms on the right-hand side of the SGLE. 
Each of them contains separate contributions from the $L$, $R$ phonons, and the
electron bath ($e$), such that ${\Pi^r}=\Pi^r_L + \Pi^r_R+\Pi^r_e$ and
$f=f_L+f_R+f_e$. The phonon self-energy $\Pi^r$ describes the time-delayed backaction of the bath
on the system due to its motion\cite{FEYNMAN1963,CALE.1983,SC.1982,JMP.2010,luprb12}. The second quantum term $f(t)$ is a random force (noise) due to the thermal, or current-induced fluctuation of the bath variables. It is characterized by the correlation matrix $\langle f_\alpha(t)
f_\alpha^T (t')\rangle = S_\alpha(t-t')$, with $\alpha=L, R, e$.  The two phonon baths ($L$ and $R$) are assumed to
be in thermal equilibrium. Their noise correlation $S_{L/R}$ is related to the
$\Pi^r_{L/R}$ through the fluctuation-dissipation theorem, $S_{L/R}(\omega) =
\left(n_B(\omega,T\right)+\frac{1}{2})\Gamma_{L/R}(\omega)$ with
$\Gamma_{L/R}(\omega) = -2 {\rm Im}\Pi^r_{L/R}(\omega)$, $n_B$ the Bose
distribution function (using atomic units, $\hbar =1$). Due to the electrical
current, the electronic bath is not in equilibrium. We define the
coupling-weighted electron-hole pair density of states
as,\cite{JMP.2010,luprb12}
\begin{widetext}
\begin{eqnarray}
\label{eq:ehdos}
\Lambda^{\alpha\beta}_{kl}(\omega)&=&2\sum_{m,n}
\langle \psi_{m}|M^k|\psi_{n}\rangle\langle \psi_{n}|M^l| \psi_{m}\rangle
(n_F(\varepsilon_n-\mu_\alpha)-n_F(\varepsilon_m-\mu_\beta))\delta(\varepsilon_n-\varepsilon_m-\omega),
\end{eqnarray}
\end{widetext}
with $n_F$ the Fermi-Dirac distribution, and $\psi_{n}$ the electron scattering state originating from the $n$-th channel of electrode $\alpha$ when there is no e-ph interaction. The noise correlation and the backaction term of the electron bath can now be written as,
\begin{eqnarray}
	&&S_e(\omega) = -2\pi\sum_{\alpha\beta}\left[n_B(\omega-(\mu_\alpha-\mu_\beta))+\frac{1}{2}\right]\Lambda^{\alpha\beta}(\omega),\nonumber\\
          \\
	&&\Pi^r_e(\omega) = -\frac{1}{2}\left( \mathcal{H}\{\Gamma_e(\omega')\}(\omega) + i \Gamma_e(\omega) \right),\\
	&&\Gamma_e(\omega)= -2\pi\sum_{\alpha\beta}\Lambda^{\alpha\beta}(\omega),
\end{eqnarray}
where $\mathcal{H}\{A\}$ is the Hilbert transform of $A$.

In the absence of electrical current, the electrons serve as an equilibrium
thermal bath, similar to phonons.
However, in the presence of current, the term ($\sim {\rm Im}\Lambda_{kl}^{RL},
k\neq l$) becomes important. It may coherently couple two vibrational modes
($kl$) inside the system leading to non-zero NC and BP forces. In
Eq.~(\ref{eq:ehdos}) we observe that these effects depend on the phase of
the electronic wavefunction, and thus the direction of electronic current.
Furthermore, the coherent coupling breaks time-reversal symmetry of the noise
correlation function, $S_e(t-t') \neq S_e(t'-t)$.
Hereafter, we denote these forces by {\em asymmetric CIF},
and focus on their role for the excess heat distribution and heat transport in
the junction.

We will consider the case where all baths are at the same temperature ($T$), and the electron bath is subject to a nonzero voltage bias ($eV=\mu_L-\mu_R$).
To look at the excess heating, we calculate the kinetic energy of atom $n$ from
its local displacement correlation function, and obtain 
\begin{eqnarray}
	E_n&=&\sum_{\sigma=x,y,z}\int_{0}^{+\infty}\!\! \omega^2 {\rm diag}\{D^r S D^a\}_{n,\sigma}(\omega) \frac{d\omega}{2\pi}.
	\label{}
\end{eqnarray}
Here $D^r$ ($D^a$) is the $eV$-dependent phonon retarded (advanced) Green's
function, $S$ is the sum of noise correlation function from all the baths, and
${\rm diag}\{A\}_{n,\sigma}$ means the diagonal matrix element of $A$,
corresponding to the $n$-th atom's $\sigma$ degrees of freedom.

To study heat transport, we calculate the phonon heat current flowing
\emph{into} the bath $L$ as the product of the velocity of the system degrees
of freedom, and the force exerted on them by bath $L$. Applying time average, using the solution of
the SGLE, we arrive at a Landauer-like expression (Sec. I, Supplemental
Materials (SM))
\begin{eqnarray}
	\label{eq:hp3}
	{J}_{L}&=& -\int_{-\infty}^{+\infty} \!\omega\, {\rm tr}\left[ \Gamma_L(\omega) D^r(\omega) \Lambda^{RL}(\omega) D^a(\omega) \right]\nonumber\\
	&&\times\left( n_B(\omega+eV)-n_B(\omega) \right)d\omega.
\end{eqnarray}
Defining the time-reversed phonon spectral function from the left bath $\tilde{\mathcal{A}}_L=D^a \Gamma_L D^r$, and similarly $\mathcal{A}_e = D^r \Lambda^{RL} D^a$, we can write the trace in Eq.~(\ref{eq:hp3}) in different forms
\begin{equation}
	\label{eq:hp3sup}
	{\rm tr}[ \Gamma_L D^r\Lambda^{RL} D^a] = {\rm tr}[ \Gamma_L\mathcal{A}_e] ={\rm tr}[ \Lambda^{RL} \tilde{\mathcal{A}}_L]. 
\end{equation}

Equations~(\ref{eq:hp3}) is analogous to the Landauer or non-equilibrium Green's function formula for electron/phonon
transport. In our present case the energy current is driven by a non-thermal
electron bath with the bias showing up in the Bose distributions and in the
coupling function, $\Lambda^{RL}$, between phonons and electrical current. The two forms in
Eq.~(\ref{eq:hp3sup}) emphasize two aspects of the problem. In the first version emphasis is on the coupling, $\Gamma_L$  of the system vibrations as described by
$\mathcal{A}_e$, to the phonons of the leads. This is a general formula, which
does not explicitly depend on the situation we are considering here, namely
that the source of energy is the non-equilibrium electron bath. This aspect is
emphasized in the second version. Here the coupling to the electrical current,
$\Lambda^{RL}$ is made explicit, and the complete phonon system including the
coupling to leads are in the function $\tilde{\cal{A}}_L$.  In both forms the
asymmetric CIF show up in the different versions of the $\cal{A}$ functions. The forces are responsible for the build up of
vibrational energy inside the junction, a fact that is present in the two 
phonon Green's functions $D^r$ and $D^a$. Apart from this effect the non-equilibrium
nature of the electron system shows up in the explicit factor $\Lambda^{RL}$
in the second version of Eq.~(\ref{eq:hp3sup}). This will develop an imaginary part which is not present in equilibrium.

Applying these formulas to a minimal model, in Sec. II of the SM, we have shown
analytically that the asymmetric CIF, especially the NC force, generate
an asymmetric phonon heat flow and energy distribution, even for a left-right
symmetric system.

\emph{First Principles calculations--}
\label{sec:auchain}
Next we turn to numerical calculation for two concrete nano-junctions. We use
SIESTA/TRANSIESTA\cite{Soler.02,BrMoOr.2002} to calculate the electronic
transport, vibrational modes, e-ph coupling employing
Ref.~\onlinecite{FrPaBr.2007}, and coupling to electrode phonons using
Ref.~\onlinecite{EnBrJa.2009}, with similar parameters.  The effect of current
on the stability of gold single atomic junctions has been studied for more than
a decade\cite{YaSa.1997,Oshima2013}. Here we first consider a symmetric single
atom gold chain between two Au(100) electrodes(Fig.~\ref{fig:AuHeat}
inset).\cite{OhKoTa98,YaBova.98.Formationandmanipulation}  We have
previously\cite{LuBrHe.2011} studied the asymmetric forces in this system
neglecting the coupling to electrode phonons.

\begin{figure}[!htbp]
\includegraphics[scale=0.6]{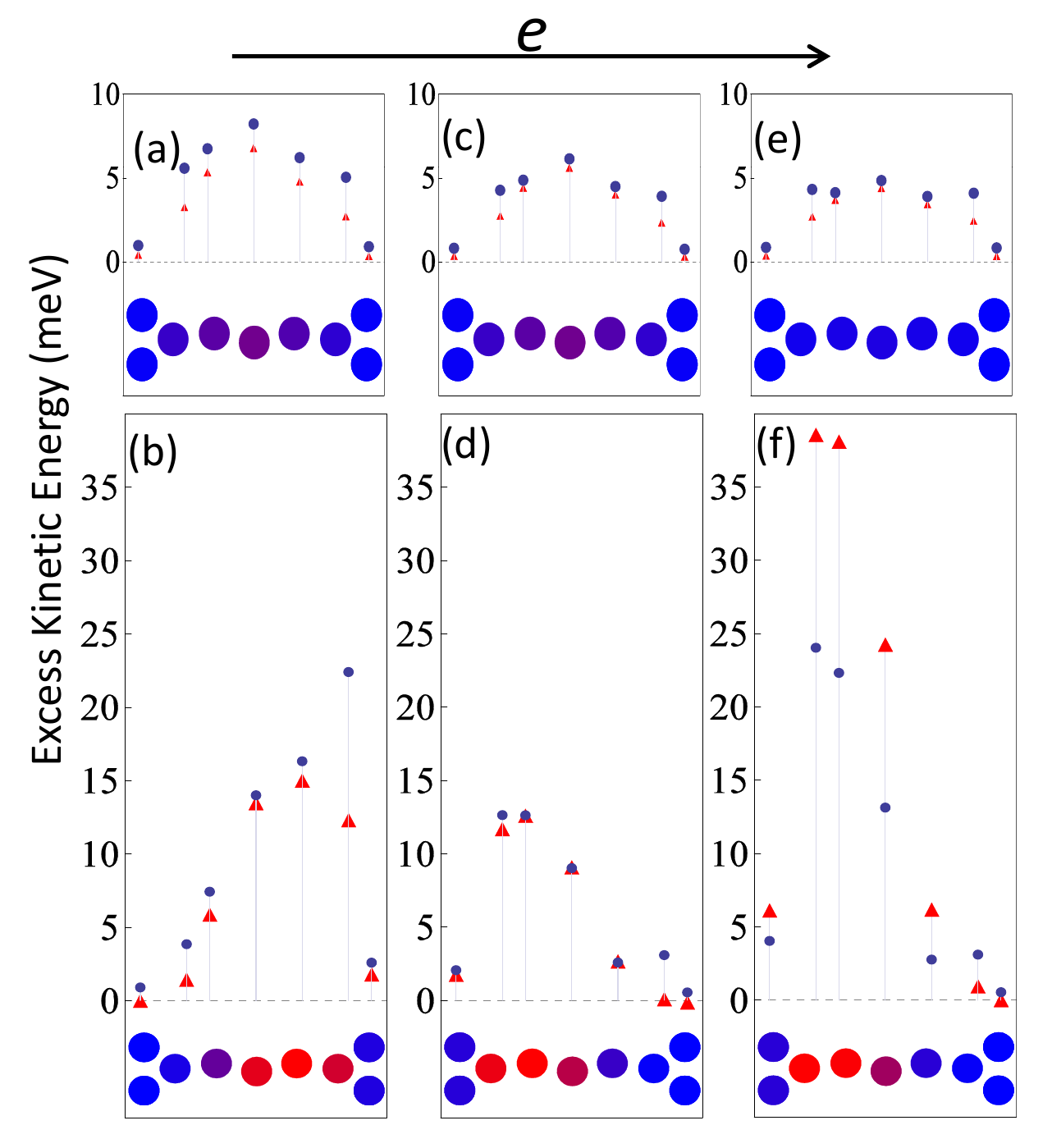}
\caption{Excess kinetic energy of each atom in a gold chain (inset of Fig.~\ref{fig:AuHeat}(a)) at $V=1.0$ V, $T=300$ K, with
(bottom) and without (top) the asymmetric CIF. The total
energy difference between the two cases is due to the non-conservative force
contribution. The blue dots and the colored plot of each atom are from the full
calculation. The asymmetric heating is qualitatively reproduced by only considering electron coupling with vibrational modes (1) and (2) in the inset of Fig.~\ref{fig:AuHeat} (a), as shown by red triangles. (a)-(b)$E_F = -0.3$ eV, (c)-(d) $E_F = 0$, and (e)-(f) $E_F=0.2$ eV. The arrow
indicates the current direction.}
\label{fig:AuT}
\end{figure}
\begin{figure}[!htbp]
\includegraphics[scale=0.8]{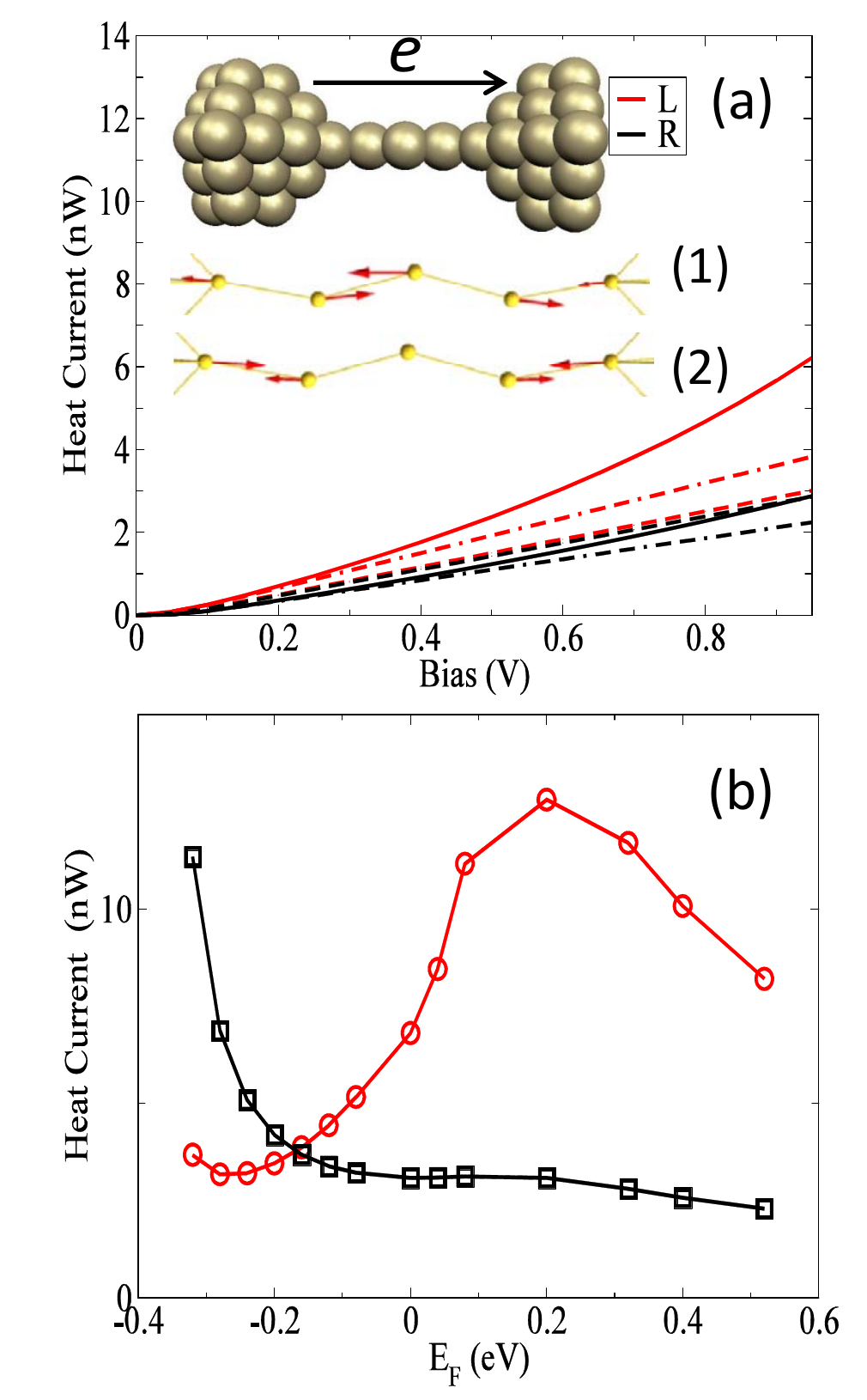}
\caption{(a) Bias dependence of the phonon heat current, going into the left
	and right phonon baths. Solid lines include the asymmetric CIF ($\sim {\rm Im}\Lambda^{RL}$), dashed lines do not, and the dash-dotted lines ignore the change of phonon spectral ($D^r/D^a$) due to NC and BP forces.  In the inset, we show the two vibrational modes that couple most strongly with the electrical current, with vibrational energy at (1)$~19$ and (2) $~18$ meV. (b) Phonon heat current going into the left (red, circle) and right (black, square) baths at $V=1$V, for different Fermi levels to illustrate the importance of the phase of the electron wavefunctions.}
\label{fig:AuHeat}
\end{figure}

Figure~\ref{fig:AuT} shows the average excess kinetic energy ($\Delta E_n =
E_n(eV)-E_n(0)$)\cite{DubiDVentra09-1,DubiDVentra09-2,Jacquet09,Jacquet12,Bergfield13,note2}
of atoms along the chain for three different Fermi level $E_F$. The structure
is almost mirror symmetric.  When we turn off the asymmetric CIF (${\rm
Im}\Lambda^{RL}=0$) as in previous studies\cite{FrBrLoJa.2004,HuChDa.2007}, the
heating profile follows this symmetry. However, once we include them, the
kinetic energy of one side becomes many times higher than that of the other.
Meanwhile, the total kinetic energy stored in the system increase
significantly.  Further analysis shows that both effects are due to the NC
force (Fig. $2$ in SM).

We now turn to the phonon heat current calculated using Eq.~(\ref{eq:hp3}),
shown in Fig~\ref{fig:AuHeat} (a). The inclusion of the asymmetric CIF drives
much larger heat current into the $L$ bath.  Intuitively, this is due to the
asymmetric energy accumulation induced by the NC force, e.g., modifying
$D^r/D^a$ in Eqs.~(\ref{eq:hp3}-\ref{eq:hp3sup}).  However, there is another
contribution at low bias.  Ignoring the bias-induced change of
$\tilde{\mathcal{A}}_L$, we get opposite heat flow into $L$ and $R$($J_L=-J_R$) due to
${\rm tr}[{\rm Im}\Lambda^{RL}{\rm Im}\tilde{\mathcal{A}}^0_L]$.  This term
drives asymmetric heat flow even in the linear response regime, contributing with a correction to the thermoelectric Peltier coefficient (Sec. I(A) of SM). In the next section, we will show that it
can be understood as asymmetric excitation of left- and right-travelling phonon
waves.

From Fig.~\ref{fig:AuT} (b)-(d) and \ref{fig:AuHeat} (b), we see that the
position of $E_F$  is controlling the direction and magnitude of the asymmetry.
According to the analysis in Sec. IV of SM, this could be due to the phase
change of the electronic wavefunction with $E_F$. Thus we expect that the direction of
electron flow is essential in the description of the atomic dynamics in the
junction, as indicated in recent experiments\cite{Schirm2013}.

\begin{figure}[!htbp]
\includegraphics[scale=0.7]{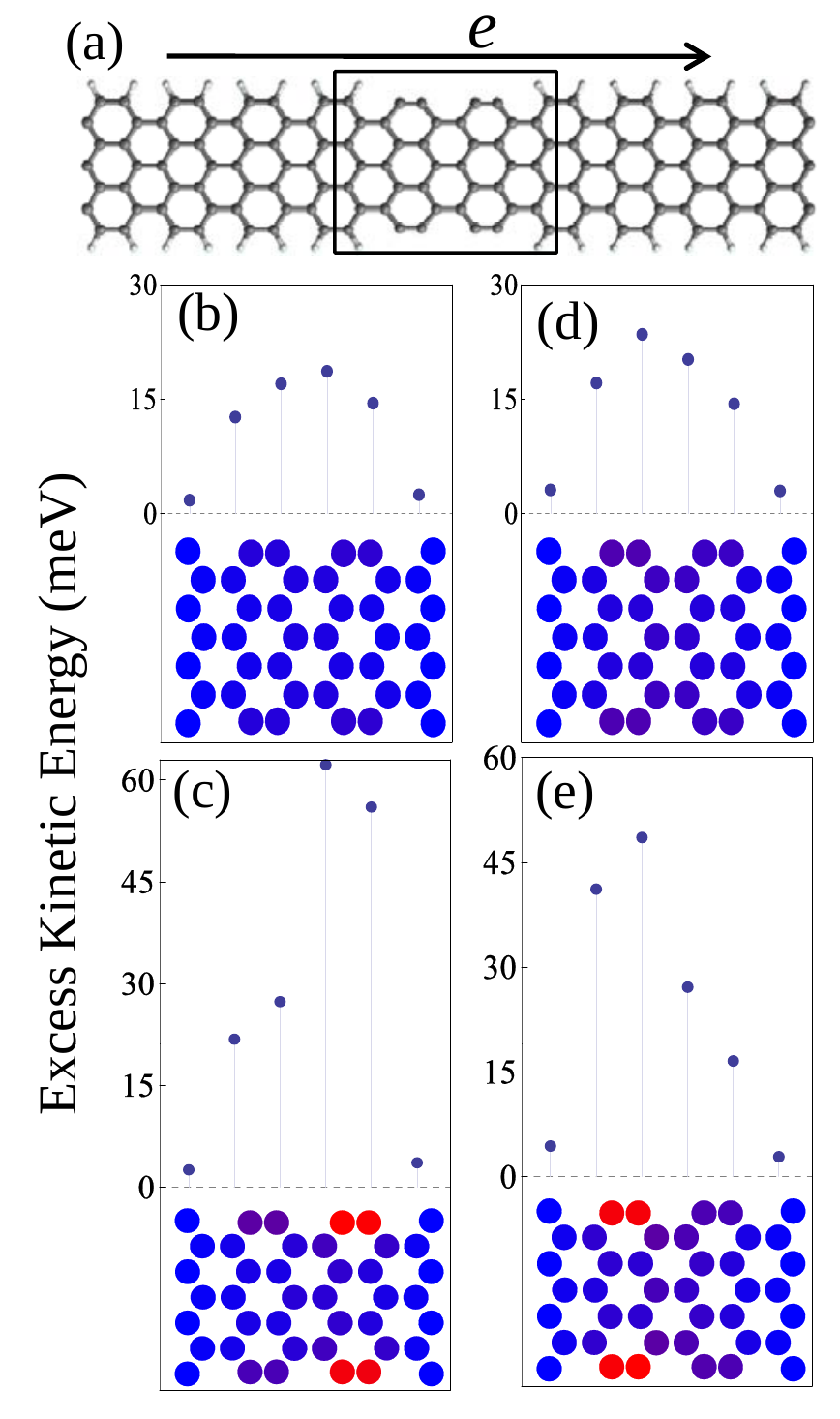}
\caption{(a) Structure of a partially passivated armchair graphene ribbon
considered. The two sides of the ribbon is hydrogen passivated except in the
device region, enclosed by the solid lines. (b)-(c) The excess kinetic energy
of each atom without and with the asymmetric CIF, at $V =
0.4$ V, $T=300$ K, $E_F=1.4$ eV. The dots show the average over atoms belonging
to each zigzag column. (d)-(e) Same with (b)-(c) with $E_F=-1.0$ eV.}
\label{fig:aGNR}
\end{figure}

The second system we consider is an armchair graphene nanoribbon (a-GNR) with
partial hydrogen passivation, shown in Fig.~\ref{fig:aGNR} (a).  This example
is inspired by experiments showing current-induced edge-reconstructions in
graphene\cite{Jia09} where the physical mechanism was attributed to Joule
heating\cite{madse10}.  In Fig.~\ref{fig:aGNR} (a), the four pairs of
unpassivated carbon dimers give rise to localized high-frequency vibrations
interacting strongly with electrical current.  Consequently, the excess energy
is mainly stored in the dimers and nearby atoms (Fig.~\ref{fig:aGNR} (b),(d)),
consistent with the experimental findings in Ref.~\onlinecite{Jia09}.
Including the asymmetric CIF leads to symmetry breaking of
the heating profile along the current direction.  Contrary to experiments on
the gold chain $E_F$ may in this case be tuned by gating. We predict the
resulting hot-spot to move from ``down-stream'' to ``up-stream'' w.r.t. the
electron current when tuning from $E_F = 1.4$ eV to $E_F = -1.0$ eV
(Fig.~\ref{fig:aGNR} (c),(e), and Fig. $3$ in SM). Thus, our calculation
further suggests that which part of the edge bonds break first may be controlled
by gating.

The dependence of the hot-spot on $E_F$
can be understood as follows (Sec. III of SM).  For a mirror-symmetric system
with electron-hole symmetry, the asymmetric heating and heat flow is absent.
When $E_F$ crosses the electron-hole symmetric point, the dominant
current-carriers contributing to inelastic transport change from electrons to
holes, or vice versa. Thus, the hot-spot moves from one side to the other.
Interestingly enough, similar effect in micrometer scale has been observed 
experimentally in graphene transistors\cite{Freitag2010,Bae2010} and
electrodes of molecular junctions\cite{Lee2013}. Here we show that it is
equally important at atomic scale, and related to the asymmetric CIF.

\emph{Scattering analysis -- }
The asymmetric heating and phonon heat flow at low bias can be qualitatively understood
from the momentum transfer between electrons and phonons. To show this, we
consider a simple 1D model with a local e-ph interaction which involve the
displacement of the $n$- and $n+1$-th atoms (junction) (Sec. IV of SM),
\begin{equation}
	\label{eq:eph2}
H_{eph} = \sum_{j\in\{n,n+1\}}-m \hat{u}_j (c^\dagger_j c_{j+1}-c^\dagger_j c_{j-1}+h.c.).
\end{equation}
For $eV>0$, the important process is the inelastic electronic
transition from the filled, left scattering states with momentum $k_L$ to the empty, right
states with $k_R$. It is straightforward to show that the emission probability of a 
right-travelling phonon with momentum $q$ is different from that of a
left-travelling mode, $-q$, due to the difference in matrix elements for the processes,  
\begin{equation}
	\Delta M_{LR} = |M^{q}_{LR}|^2-|M^{-q}_{LR}|^2 \sim \sin (q) \sin (k_L-k_R).
	\label{eq:dq}
\end{equation}
Consequently, the left- and right-travelling steady state phonon populations
become different, resulting in asymmetric heat flow.

In conclusion, we have presented a theory showing that CIF
in nano-junctions lead to asymmetric distributions and transport of the excess
heat.  We derived a Landauer-like formula for the excess heat transport.
Employing first-principles calculations, we demonstrate that the size of the
asymmetry can be crucial for current-induced processes at the atomic scale.

We thank T. N. Todorov, D. Dundas, and T. Markussen for discussions and the
Danish Center for Scientific Computing (DCSC) for computer resources.
This work is supported by the Lundbeck Foundation (R49-A5454), National Natural
Science Foundation of China (Grants No. 11304107, 61371015), and the
Fundamental Research Funds for the Central Universities (HUST:2013TS032). 
\bibliography{Langevin}

\begin{thebibliography}{50}
\bibitem[{not({\natexlab{a}})}]{note}
\bibinfo{note}{We use phonons and vibrations interchangably, although, strictly
  speaking, phonons are defined only in systems with translational invariance.}
\bibitem[{not({\natexlab{b}})}]{note2}
\bibinfo{note}{Another way of quantifying the heating is to use the local
  temperature defined in some way. We tried to use the method in Refs. [40-44].
  The result is shown in Fig. 3 of the SM. The overall heating profile agrees
  with Fig. 1.}
\end{thebibliography}

\clearpage
\onecolumngrid
\section*{SUPPLEMENTAL MATERIALS}
\section{Derivation of the phonon heat current Eq.~(8)}
We start from the semi-classical generalized Langevin equation (SGLE) (Eq.~(2)
in the main text). To study the
energy transport, we look at the energy increase of the system per unit time
\begin{eqnarray}
	\dot E_{S}(t) &=& \frac{d}{dt}\left(\frac{1}{2}\dot U^T \dot U + \frac{1}{2}U^TKU\right) \nonumber \\
		&=& -\dot U^T\left(\sum_\alpha \int^{t}_{-\infty} \Pi^r_{\alpha}(t-t')U(t') dt'- f_{\alpha}(t)  \right), \quad \alpha=L,R,e. 
	\label{eq:pows}
\end{eqnarray}
Note that the system includes only the atomic degrees of freedom.  We can
define the energy current flowing \emph{into} the bath $\alpha$ from the system
\begin{equation}
	J_\alpha (t) \equiv \dot U^T\left(\int^{t}_{-\infty} \Pi^r_{\alpha}(t-t')U(t') dt' - f_{\alpha}(t)\right).
	\label{eq:ia}
\end{equation}
At steady state we have
\begin{equation}
	-\dot E_{s} \equiv J_e+J_{L}+J_R\equiv J_e+J_{ph} = 0.
	\label{eq:bal}
\end{equation}
We can write the expression for the average energy current in the frequency domain,
\begin{eqnarray}
	J_\alpha &\equiv& \lim_{T\to +\infty}\frac{1}{T}\int_{0}^T \left\langle \dot U^T(t)\left(\int^{t} \Pi^r_{\alpha}(t-t')U(t') dt'- f_{\alpha}  \right)\right\rangle dt\nonumber\\
	&=& \lim_{T\to +\infty}\frac{1}{T}\int \frac{d\omega}{2\pi} \langle \dot U^\dagger(\omega)\left({\Pi^r}_{\alpha}(\omega)U(\omega)-f_{\alpha}(\omega)\right)\rangle
	\label{eq:pel0}
\end{eqnarray}
Now we use the solution of the Langevin equation 
\begin{eqnarray}
	U(\omega) &=& -D^r(\omega)f(\omega),\\
	D^r(\omega)&=&\left[ \omega^2-K-\Pi^r(\omega) \right]^{-1},\\
	\Pi^r(\omega)&=&\Pi_L(\omega)+\Pi_R(\omega)+\Pi_e(\omega),
\end{eqnarray}
and the noise correlation function
\begin{eqnarray}
	\langle f_\alpha(\omega)f_\alpha(\omega')\rangle &=& \delta(\omega+\omega') S_\alpha(\omega),\\
	S(\omega)&=&S_L(\omega)+S_R(\omega)+S_e(\omega),
	\label{eq:slan2}
\end{eqnarray}
to get ($\hbar=1$)
\begin{eqnarray}
	\label{eq:pel1}
	J_\alpha&=& i\int_{-\infty}^{+\infty} \frac{d\omega}{2\pi} \omega {\rm Tr}\left[\Pi^r_{\alpha}(\omega) D^r(\omega)  S(\omega)  D^a(\omega)  + S_\alpha(\omega)  D^a(\omega) \right]\\
	&=& i\int_{0}^{+\infty} \frac{d\omega}{2\pi} \omega {\rm Tr}\left[\Pi^r_{\alpha}(\omega) D^r(\omega)  S(\omega)  D^a(\omega)  + S_\alpha(\omega)  D^a(\omega)\right.\\
	&&-\left. \Pi^r_{\alpha}(-\omega) D^r(-\omega)  S(-\omega)  D^a(-\omega)  - S_\alpha(-\omega)  D^a(-\omega)\right].\label{eq:treq}
\end{eqnarray}

The two phonon baths ($L$ and $R$) are assumed to be in thermal equilibrium. Their noise correlation $S_{L/R}$ is related to the $\Pi^r_{L/R}$ through the fluctuation-dissipation theorem, $S_{L/R}(\omega) = \left(n_B(\omega,T\right)+\frac{1}{2})\Gamma_{L/R}(\omega)$ with $\Gamma_{L/R}(\omega) = -2 {\rm Im}\Pi^r_{L/R}(\omega)$, $n_B$ the Bose distribution function (using atomic units, $\hbar =1$). The noise correlation of the electron bath is given by Eqs.~(3)-(4) in the main text.
Using the following properties
\begin{eqnarray}
	&&(D^r)^\dagger(\omega) = D^a(\omega),
	(\Pi^r)^\dagger(\omega) = \Pi^a(\omega),
	\Gamma(\omega) = i(\Pi^r(\omega)-\Pi^a(\omega)),\\
	&&S^\dagger(\omega) = S(\omega),
	S(-\omega) = S^*(\omega),
	D^r(-\omega)=(D^r)^*(\omega),
	D^a(-\omega)=(D^a)^*(\omega),
	\label{eq:rel}
\end{eqnarray}
and taking transpose of Eq.~(\ref{eq:treq}), we get a compact form
\begin{eqnarray}
	J_\alpha &=& \int_{0}^{+\infty} \frac{d\omega}{2\pi} \omega {\rm Tr}\left[\Gamma_{\alpha}(\omega) D^r(\omega)  S_{\bar{\alpha}}(\omega)  D^a(\omega)  - S_\alpha(\omega)D^r(\omega)\Gamma_{\bar{\alpha}}(\omega)D^a(\omega)\right].
	\label{eq:fib2}
\end{eqnarray}
This result has a clear physical meaning. Here, $\Gamma_\alpha$ characterizes coupling of the $\alpha$ bath to the system, and $S_{\bar{\alpha}}$ represents the energy source from all other baths.  The first term in the trace
represents energy flow into bath $\alpha$ from other baths; while
the second one represents the opposite process.

\subsection{Current-induced phonon heat transport}
\label{sec:dcur}
Now suppose all the baths are at the same temperature ($T$), but the electron
bath is subject to a nonzero bias ($eV$). The energy current
injecting into the phonon bath ($L$) is
\begin{eqnarray}
	\label{eq:fib31}
	J_L^{} &=&\int_{0}^{+\infty} \frac{d\omega}{2\pi} \omega {\rm Tr}\left[\Gamma_{L}(\omega) D^r(\omega)  S_{\bar{\alpha}}(\omega)  D^a(\omega)  - S_L(\omega)D^r(\omega)\Gamma_{\bar{\alpha}}(\omega)D^a(\omega)\right]\\
	\label{eq:fib32}
	&=& \int_{0}^{+\infty} \frac{d\omega}{2\pi} \omega {\rm Tr}\left[\Gamma_{L}(\omega) D^r(\omega)  S_e(\omega)  D^a(\omega)  - S_L(\omega)D^r(\omega)\Gamma_e(\omega)D^a(\omega)\right].
\end{eqnarray}
To go from Eq.~(\ref{eq:fib31}) to (\ref{eq:fib32}), we notice that the energy flow from $L$ to $R$ is the same as that from $R$ to $L$, since they are at the same
temperature. Thus, the only energy source is the electron bath. Using Eqs.~(3-6) in the main text, the heat current now reads
\begin{eqnarray}
	{J}_L&=&-\sum_{\alpha\neq\beta}\int_0^{+\infty} d\omega \omega {\rm Tr}\left[ \Lambda^{\alpha\beta}(\omega) D^a(\omega) \Gamma_L(\omega) D^r(\omega)\right]
	\left( n_B(\omega-(\mu_\alpha-\mu_\beta))-n_B(\omega)\right)\\
	&=&-\int_0^{+\infty} d\omega \omega {\rm Tr}\left[ \Lambda^{RL}(\omega) D^a(\omega) \Gamma_L(\omega) D^r(\omega)\right]
	\left( n_B(\omega+eV)-n_B(\omega)\right).
	\label{eq:i}
\end{eqnarray}
Define the time-reversed phonon spectral function $\tilde{\mathcal{A}}_L(\omega) = D^a(\omega)\Gamma_L(\omega)D^r(\omega)$, we can write it in other equivalent forms
\begin{eqnarray}
	\label{eq:ii0}
	{J}_L&=&-\int_{-\infty}^{+\infty} d\omega \omega {\rm Tr}\left[ \Lambda^{LR}(\omega) \tilde{\mathcal{A}}_L(\omega)\right]\left( n_B(\omega-eV)-n_B(\omega)\right)\\
\label{eq:ii2}
&=&-\int_{-\infty}^{+\infty} d\omega \omega {\rm Tr}\left[ \Lambda^{RL}(\omega) \tilde{\mathcal{A}}_L(\omega)\right]\left( n_B(\omega+eV)-n_B(\omega)\right).
\end{eqnarray}
Similar equation holds for ${J}_{R}$
\begin{eqnarray}
	\label{eq:ir0}
	{J}_{R}&=&-\int_{-\infty}^{+\infty} d\omega \;\omega {\rm Tr}\left[ \Lambda^{RL}(\omega) \tilde{\mathcal{A}}_R(\omega)\right]\left( n_B(\omega+eV)-n_B(\omega)\right)\\
	\label{eq:ir}
	&=&-\int_{-\infty}^{+\infty} d\omega \;\omega {\rm Tr}\left[ \Lambda^{LR}(\omega) \tilde{\mathcal{A}}_R(\omega)\right]\left( n_B(\omega-eV)-n_B(\omega)\right).
\end{eqnarray}

Let's look at the low bias situation.
We ignore the change of $\tilde{\mathcal{A}}_{L/R}$, and
replace it with $\tilde{\mathcal{A}}_L^0$, the counterpart of
$\tilde{\mathcal{A}}_L$ without coupling to electrons. The asymmetric
current-induced forces ($\sim {\rm Im}\Lambda^{RL}$) drive a heat current
\begin{eqnarray}
	\label{eq:hp5}
	{J}^0_{L,p}&=&\frac{1}{2}\int_{-\infty}^{+\infty} d\omega \;\omega {\rm Tr}\left[{\rm Im} \Lambda^{RL}(\omega) {\rm Im}\tilde{\mathcal{A}}^0_L(\omega)\right]\left( \coth\left(\frac{\omega+eV}{2k_BT}\right)-\coth\left(\frac{\omega}{2k_BT}\right) \right),
\end{eqnarray}
The expression for $J_{R,p}^0$ is obtained by replacing
$\tilde{\mathcal{A}}_L^0$ with $\tilde{\mathcal{A}}_R^0$.  From ${\rm
Im}\tilde{\mathcal{A}}^0_L+{\rm Im}\tilde{\mathcal{A}}^0_R=0$, we get
$J^0_{L,p}=-J^0_{R,p}$. That is, the heat flowing into bath $L$ and $R$ is
opposite.  This makes $J_{L}\neq J_{R}$, even for a symmetric structure. Furthermore, in the linear response regime, considering thermoelectric transport, from Eq.~(\ref{eq:hp5}) we get a correction to the Peltier coefficient due to electron-phonon interaction: The applied bias drives a phonon heat current from one phonon bath to the other.

From the derivation of Eq.~(\ref{eq:hp5}), and~(\ref{eq:pel1}),
we observe that, the first term with $\omega \coth((\omega+eV)/(2k_BT))$ in
$J^0_{L,p}$ is contributed by the fluctuating force in the SGLE, while the
second term with $\omega \coth(\omega/(2k_BT))$ is from the deterministic NC
force. If the bias $|eV|$ is much higher than the phonon frequency, the
contribution from NC force dominates.  This can be seen from the symmetry of
the functions, as follows: for high enough bias, $\omega
\coth((\omega+eV)/(2k_BT))$ is close to be odd in $\omega$, e.g., ignoring
$\omega$ in the coth function. But $\omega  \coth(\omega/(2k_BT))$ is even in
$\omega$. Meanwhile, the trace in Eq.~(\ref{eq:hp5}) can be approximated by an
even function for small $\omega$. Thus, the contribution of the NC force
dominates. The above analysis based on Eq.~(\ref{eq:hp5}) is correct to the 2nd
order in $M$.  Going beyond the 2nd order, we notice that in
Eqs.~(\ref{eq:ii0}-\ref{eq:ir}), the deterministic NC and BP force modifies the
phonon spectral function $\tilde{\mathcal{A}}_{L/R}$, while the fluctuating
force has no effect on it. Altogether, we conclude that, the asymmetric noise
has a negligible contribution to the asymmetric heat flow.

\section{Minimal model} \label{sec:mini} We now consider a minimal model with
two atomic vibrations. In additional to electrons, they
couple symmetrically to the left and right phonon bath, respectively. This
gives rise to lifetime broadening of $\gamma_e$ and $\gamma_{ph}$, respectively. 
The phonon Green's function is written as 
\begin{equation} 
	D^r(\omega) =  \frac{1}{N} \left( \begin{array}{cc}
	\Omega & -\omega_1^2-a-i b \omega\\ -\omega_1^2+a+i b \omega&\Omega
\end{array} \right).  \label{eq:mydr} 
\end{equation} 
Here, $\Omega=\omega^2 - \omega_0^2 +i\gamma_t\omega$, $\gamma_t=\gamma_{e}+\gamma_{ph}$,
$N=\Omega^2-(\omega_1^2+a+i b\omega)(\omega_1^2-a-i b\omega)$, $a$ and $b$ are
due to NC and BP forces, respectively. Finally, $\omega_0$ is the atomic
vibration frequency, and $\omega_1$ characterizes the coupling between the two
sites. We have ignored a term $\sim -i\gamma'_e\omega$ in the off-diagonals of
$D^r(\omega)$.  The advanced Green's function is $D^a= (D^r)^\dagger$. We also
have 
\begin{equation} 
	\tilde{\Pi}_L(\omega) =-2i \gamma_{ph}\omega \left(
	\begin{array}{cc} 1 & 0\\ 0&0 \end{array} \right).  \label{}
\end{equation} 
From these, we get the time-reversed phonon left spectral function 
	\begin{equation} \tilde{\mathcal{A}}_L(\omega) = \frac{2\omega
		\gamma_{ph}}{|N|^2} \left( \begin{array}{cc} |\Omega|^2 & -(a+i
			b \omega+\omega_1^2) \Omega^*\\
			-(a-ib\omega+\omega_1^2)\Omega&\omega_1^4+a^2+b^2\omega^2+2a\omega_1^2
		\end{array} \right), \label{} \end{equation}
\subsection{Heat current}
To calculate the heat current, we assume 
\begin{eqnarray} \label{eq:glam} 
\Lambda^{\alpha\beta}(\omega) &\approx&-2(\omega-(\mu_\alpha-\mu_\beta)) \left( \begin{array}{cc}
\lambda^{\alpha\beta}_{1} & \lambda^{\alpha\beta}_{2}+i
\lambda^{\alpha\beta}_3\\ \lambda^{\alpha\beta}_2-i
\lambda^{\alpha\beta}_3&\lambda^{\alpha\beta}_1 \end{array}
\right).   
\end{eqnarray} 
This means we ignore the energy dependence of the electronic properties within the bias window.
We can now evaluate the trace in Eq.~(\ref{eq:ii2}), 
\begin{eqnarray}
	\label{eq:imtr0} {\rm Tr}\left[ {\rm Im}\Lambda^{RL} {\rm
	Im}\tilde{\mathcal{A}}_L \right]
	&=&-\frac{8}{|N|^2}\omega^2(\omega+eV)\gamma_{ph}\left[\frac{1}{\omega_c}eV\lambda_3^{2}(\omega_0^2-\omega^2)+eV\lambda_3^{2}\gamma_{t}\underbrace{-\lambda_3\omega_1^2\gamma_{t}}\right],\\
	\label{eq:retr0} {\rm Tr}\left[ {\rm Re}\Lambda^{RL}{\rm
	Re}\tilde{\mathcal{A}}_L \right]&=&
	-\frac{4}{|N|^2}\omega(\omega+eV)\gamma_{ph}\left[|\Omega|^2\lambda_{1}^{}+\lambda_{1}^{}\left(
	\omega_1^4+eV^2{\lambda_3^{}}^2\left(1+\frac{\omega^2}{\omega_c^2}\right)\right)\right.\nonumber\\
	&-&\left.2\lambda_{2}^{}\omega_1^2(\omega^2-\omega_0^2)+\underbrace{2eV\lambda^{}_3\left[\lambda_{2}^{}\left(
	\frac{\omega^2}{\omega_c}\gamma_t+(\omega^2-\omega_0^2)
	\right)-\lambda_1^{}\omega_1^2\right]}\right].  \end{eqnarray} 
We have drooped the $RL$ superscript in $\lambda$s for notational simplicity,
and used $a =-eV \lambda^{RL}_3$, $b=a/\omega_c$. Here $\omega_c$ is on the
order of the electron bandwidth. Substituting back into Eq.~(\ref{eq:ii2}), we find that those terms in the curly brackets of
Eqs.~(\ref{eq:imtr0}) and (\ref{eq:retr0}), due to the asymmetric
current-induced forces ($\sim \lambda_3$), induce asymmetric
heat flow (odd in $eV$) to the left and right phonon bath.

\subsection{Average kinetic energy}
We now calculate the average kinetic energy difference between the two atomic
sites.  
If we take a general noise correlation for the electronic bath
\begin{equation}
	S_e(\omega,eV) = \sum_{\alpha,\beta=\{L,R\}} g^{\alpha\beta}(\omega)
	\left(
	\begin{array}{cc}
		\lambda^{\alpha\beta}_{1} & \lambda^{\alpha\beta}_{2}+i \lambda^{\alpha\beta}_3\\
		\lambda^{\alpha\beta}_2-i \lambda^{\alpha\beta}_3&\lambda^{\alpha\beta}_1
	\end{array}
	\right),
	\label{}
\end{equation}
with
\begin{equation}
	g^{\alpha\beta} (\omega) = 2\pi(\omega -(\mu_\alpha-\mu_\beta))\coth \left( \frac{\omega -(\mu_\alpha-\mu_\beta)}{2k_BT} \right).
	\label{}
\end{equation}
Using Eq.~(6) in the main text, we get
\begin{eqnarray}
	\Delta E^e = \sum_{\alpha,\beta={L,R}}\int \frac{\omega^2 g^{\alpha\beta}(\omega)}{|N|^2}\left[\lambda^{\alpha\beta}_2(a(\omega^2-\omega_0^2)+b\omega^2\gamma_t) -a\lambda^{\alpha\beta}_1\omega_1^2-\lambda^{\alpha\beta}_3 \gamma_t \omega\omega_1^2 \right] \frac{d\omega}{2\pi}.
	\label{}
\end{eqnarray}
We look at the nonequilibrium contribution first $S_e^{non}(\omega)=S_e(\omega,eV)-S_e(\omega,0)$. For $eV \gg \omega_0$,
similar arguments to Sec.~\ref{sec:dcur} show that the main contribution comes
from the real part of the two terms with $\alpha\neq\beta$, and the asymmetric noise is negligible,
\begin{equation}
	S^{non}_e(\omega) \sim 2\pi|eV| 
	\left(
	\begin{array}{cc}
		\lambda^{RL}_{1} & \lambda^{RL}_{2}\\
		\lambda^{RL}_2&\lambda^{RL}_1
	\end{array}
	\right),\quad eV\gg \omega_0.
	\label{}
\end{equation}
Consequently, we get
\begin{eqnarray}
	\Delta E^{non} \approx \int \frac{4\pi\omega^2eV \lambda_3^{RL}|eV|}{|N|^2}\left[ \lambda_2^{RL}\left( \omega^2-\omega_0^2+\frac{\omega^2\gamma_t}{\omega_c} \right)-\lambda_1^{RL}\omega_1^2 \right] \frac{d\omega}{2\pi}.
	\label{}
\end{eqnarray}
The BP force contribution is negligible if $\omega_c$ is the largest energy scale of the problem. If we further ignore $\lambda_2^{RL}$ to be consistent with
Eq.~(\ref{eq:mydr}), we get
\begin{eqnarray}
	\Delta E^{non} \approx -4\pi eV |eV|\lambda_3^{RL}\lambda_1^{RL}\omega_1^2 \int \frac{\omega^2 }{|N|^2}  \frac{d\omega}{2\pi}.
	\label{}
\end{eqnarray}
For the equilibrium part, including contribution from phonon baths, we get
\begin{eqnarray}
	\Delta E^{equ} \approx -eV\lambda_3^{RL} (\gamma_{ph}+\gamma_e)\omega_1^2 \int \frac{\omega^3}{|N|^2} \coth\left( \frac{\omega}{2k_BT} \right)\frac{d\omega}{2\pi},
	\label{eq:eequ}
\end{eqnarray}
The total difference $\Delta E^e = \Delta E^{non} +\Delta E^{equ}$. We see that
$\Delta E^{e} = 0$ if  $\lambda_3^{RL} = 0$. Thus,  the asymmetric current-induced forces generate asymmetric energy distribution, with the NC force contributes predominantly. The asymmetry is enhanced by coupling to phonon
baths ($\gamma_{ph}$ in Eq.~(\ref{eq:eequ})).

\section{electron-hole symmetry}
Assuming symmetrical voltage drop across the conductor, we define the zero
energy as the equilibrium Fermi level. The left and right chemical potential are
at $eV/2$ and $-eV/2$, respectively.  The $\Lambda$-function now reads 
\begin{eqnarray}
	\Lambda^{\alpha\beta}_{kl}(\omega,eV) &=& 2\sum_{m,n}\langle \psi_m|M^k|\psi_n\rangle \langle \psi_n|M^l|\psi_m\rangle 
	\left[ n_F\left(\varepsilon_n-s_\alpha eV/2\right)-n_F\left( \varepsilon_m-s_\beta eV/2\right) \right]\delta\left( \varepsilon_n-\varepsilon_m-\omega \right),
	\label{}
\end{eqnarray}
where we have written explicitly its $eV$ dependence, and $s_L=1$, $s_R=-1$. It has the following properties:
\begin{eqnarray}
	\label{eq:lams1}
	\Lambda^{\alpha\beta}_{kl}(\omega, eV) = {\Lambda_{lk}^{\alpha\beta}}^*(\omega, eV),\quad
	\label{eq:lams2}
	\Lambda^{\alpha\beta}_{kl}(\omega, eV) = -{\Lambda_{lk}^{\beta\alpha}}(-\omega, eV).
\end{eqnarray}
For the convenience of further analysis, we now use
\begin{equation}
	A_\alpha(\varepsilon) = 2\pi\sum_n |\psi_n\rangle \delta(\varepsilon-\varepsilon_n) \langle \psi_n|,
	\label{}
\end{equation}
to write it as
\begin{eqnarray}
	\label{eq:lameh}
	\Lambda^{\alpha\beta}_{kl}(\omega,eV) &=& 2\int \frac{d\varepsilon}{2\pi} \int \frac{d\varepsilon'}{2\pi} {\rm Tr}\left[ M^k A_\alpha(\varepsilon) M^l A_\beta(\varepsilon') \right] 
	\left[ n_F\left(\varepsilon-s_\alpha eV/2\right)-n_F\left( \varepsilon'-s_\beta eV/2\right) \right]\delta\left( \varepsilon-\varepsilon'-\omega \right),
\end{eqnarray}
Note that the spectral function $A_\alpha(\varepsilon)$ is Hermitian. If
we use a real-space basis set, it is a complex matrix. We define the
system has electron-hole symmetry if 
\begin{equation}
	\label{}
	{\rm Re}A_\alpha(\varepsilon) ={\rm Re}A_\alpha(-\varepsilon), \quad \rm{or} \quad
	\label{eq:ehsym} A_\alpha(\varepsilon) = A^*_\alpha(-\varepsilon),
\end{equation}
The two conditions are equivalent since ${\rm Re}A_\alpha(\varepsilon)$ is
related to ${\rm Im}A_\alpha(\varepsilon)$ by Hilbert transform, which changes
their symmetry with respect to $\varepsilon$.
Using Eq.~(\ref{eq:ehsym}) in (\ref{eq:lameh}), together with Eq.~(\ref{eq:lams1}), we find that 
\begin{equation}
	\Lambda^{LR}(\omega,eV) = \Lambda^{RL}(\omega,-eV).
	\label{}
\end{equation}
Here, we have further assumed that the electron-phonon interaction matrix is
real.  This is a reasonable assumption, if we ignore the bias-dependence of the
electronic Hamiltonian, and consider Cartesian phonon index, without external
magnetic field.  Substituting it into Eqs.~(\ref{eq:ii0}-\ref{eq:ii2}) and
(\ref{eq:ir0}-\ref{eq:ir}), we find that 
\begin{equation}
	J_\alpha(eV) = J_\alpha(-eV), \quad \alpha=L, R.
	\label{}
\end{equation}
We reach the conclusion that the heat flow into $L$ and $R$ are the same if the system has
electron-hole symmetry and there is a symmetrical voltage drop across the conductor.

\section{Scattering analysis}
The asymmetric heating and heat flow at low bias can be qualitatively
understood as momentum transfer between electrons and phonons. To show this, we
consider a simple one-dimensional (1D) model.  The electronic subsystem is
described by a nearest neighbour tight-binding Hamiltonian, and the phonon
subsystem by a harmonic oscillator model. To simplify the analysis, we assume
that the tight-binding hopping parameter and the spring constant  between all
the nearest sites are the same. But the analysis can be easily extended to more
general case, where our conclusion in this section still holds. The electron
and phonon states are described by scattering waves originating from $L$ and
$R$.  We introduce a local e-ph interaction on two atomic sites $n$ and $n+1$
(junction), that is, the displacement of the $n$- and $n+1$-th atoms modifies
the electronic hopping elements nearby linearly, e.g., for phonon mode $q$,
\begin{equation}
	M^q \sim \left( \begin{array}{cccc}
		0&1&0&0\\
		1&0&-1+e^{iq}&0\\
		0&-1+e^{iq}&0&-e^{iq}\\
		0&0&-e^{iq}&0\\
	\end{array} \right)\,.
	\label{eq:ephq}
\end{equation}

For positive bias $eV>0$, the
main process contributing to phonon emission is the inelastic electronic
transition from the filled, left scattering states $\psi_L$ to the empty, right
states $\psi_R$. The transition rate is proportional to the modulus square of
the matrix element,
\begin{eqnarray}
	|M^q_{LR}|^2 = |M^{-q}_{RL}|^2 \sim  \cos^2\frac{1}{2}(q-k_L+k_R).
	\label{eq:mes}
\end{eqnarray}
The emission probability of a right-travelling phonon mode $q$ ($q>0$) is different from that of a
left-travelling mode, $-q$.  The difference is
\begin{equation}
	\Delta M_{LR} = |M^{q}_{LR}|^2-|M^{-q}_{LR}|^2 \sim \sin (q) \sin (k_L-k_R),
	\label{eq:dq}
\end{equation}
and as a result, the left and right-travelling steady state phonon populations becomes different. 
The difference changes sign upon changing the current direction which reveal the importance of electron momentum.

\begin{figure}[!htbp]
\includegraphics[scale=0.4]{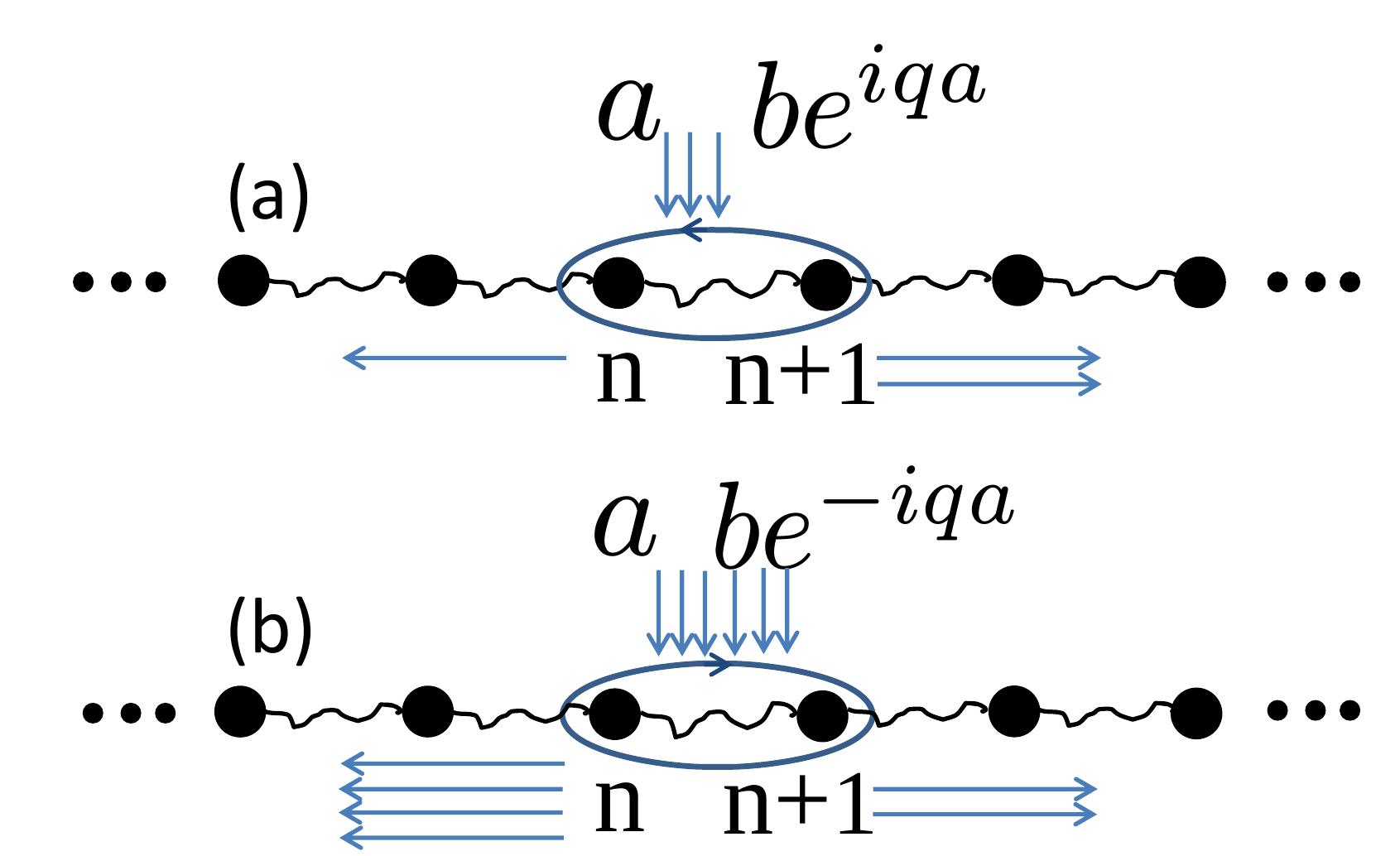}
\caption{The motion of atom $n$ and $n+1$ has a phase shift of $\pm q$. Within
the configuration space of $u_n$ and $u_{n+1}$, the two situations correspond to elliptical
motion in opposite directions. The excitation probabilities of these two modes differs.
This results in (1) an asymmetric heat flow to the left and right phonon bath, (2)
polarization of the motion within configuration space $(u_n,u_{n+1})$.}
\label{fig:flow}
\end{figure}

Next, we use the retarded phonon Green's function to consider the response, $|r\rangle$, of the phonon 
system to the asymmetric excitation, $|s\rangle \sim ( \begin{array}{cccccc} \cdots&0&1&e^{iq}&0&\cdots  \end{array} )^T$, at $n$ and $n+1$,
and find
\begin{eqnarray}
	|\langle m|r\rangle|^2 &\sim& |\langle m|D^r_0 |s\rangle|^2 \sim \left\{\begin{array}{rl}
		\cos^2\frac{q+|q|}{2}, & m\ll n\\
		\cos^2\frac{q-|q|}{2},& m\gg n
	\end{array}\right.
	\label{eq:re}
\end{eqnarray}
where obviously the response differs at the left and right side of
the perturbation (Fig.~\ref{fig:flow}).

We conclude that the applied bias breaks the population balance
between left and right electron scattering states. Consequently, electrons
excite the left and right travelling phonon states differently resulting in transfer of 
both energy and momentum to the phonons. The momentum transfer generates a
different phonon energy flux to the left and right for the spatially symmetric
system under bias.  A schematic diagram of these processes are shown in
Fig.~\ref{fig:flow}. If we turn on the e-ph interaction at all the sites, the interaction
matrix becomes $M^q_{LR} \sim \delta(k_L-k_R-q-2N\pi )$.  The asymmetric
phonon excitation reduces to the rule of crystal momentum conservation in the
periodic structure.

To make connection with the current-induced NC and
BP force, in Fig.~\ref{fig:flow} we illustrate the orbital of the two phonon
excitation within the configuration space of $(u_n,u_{n+1})$. They are
elliptical and related by time-reversal.  From this point of view, the
current-induced NC and BP forces polarize the atomic orbital motion, and generate a net angular
momentum. The heat flow into the two electrodes becomes
different due to this elliptical polarization.

Finally, it is instructive to compare the scattering analysis against the Langevin
approach. In fact, one can show that 
\begin{equation}
	{\rm Im}\langle \psi_{L}|M^{n}|\psi_{R}\rangle\langle \psi_{R}|M^{n+1}| \psi_{L}\rangle \sim \sin (k_L-k_R),
	\label{}
\end{equation}
and 
\begin{equation}
	({\rm Im}\tilde{\mathcal{A}}^0_L)_{n+1,n}(\omega_q) \sim \sin(q).
	\label{}
\end{equation}
So, comparing Eqs.~(\ref{eq:hp5}) and (\ref{eq:dq}), we can
see that the asymmetric heat flow can indeed be understood as
a result of asymmetric excitation of left- and right-travelling phonon waves.

\section{Supporting figures from the first-principles calculation}
\begin{figure}[!htbp]
\includegraphics[scale=0.6]{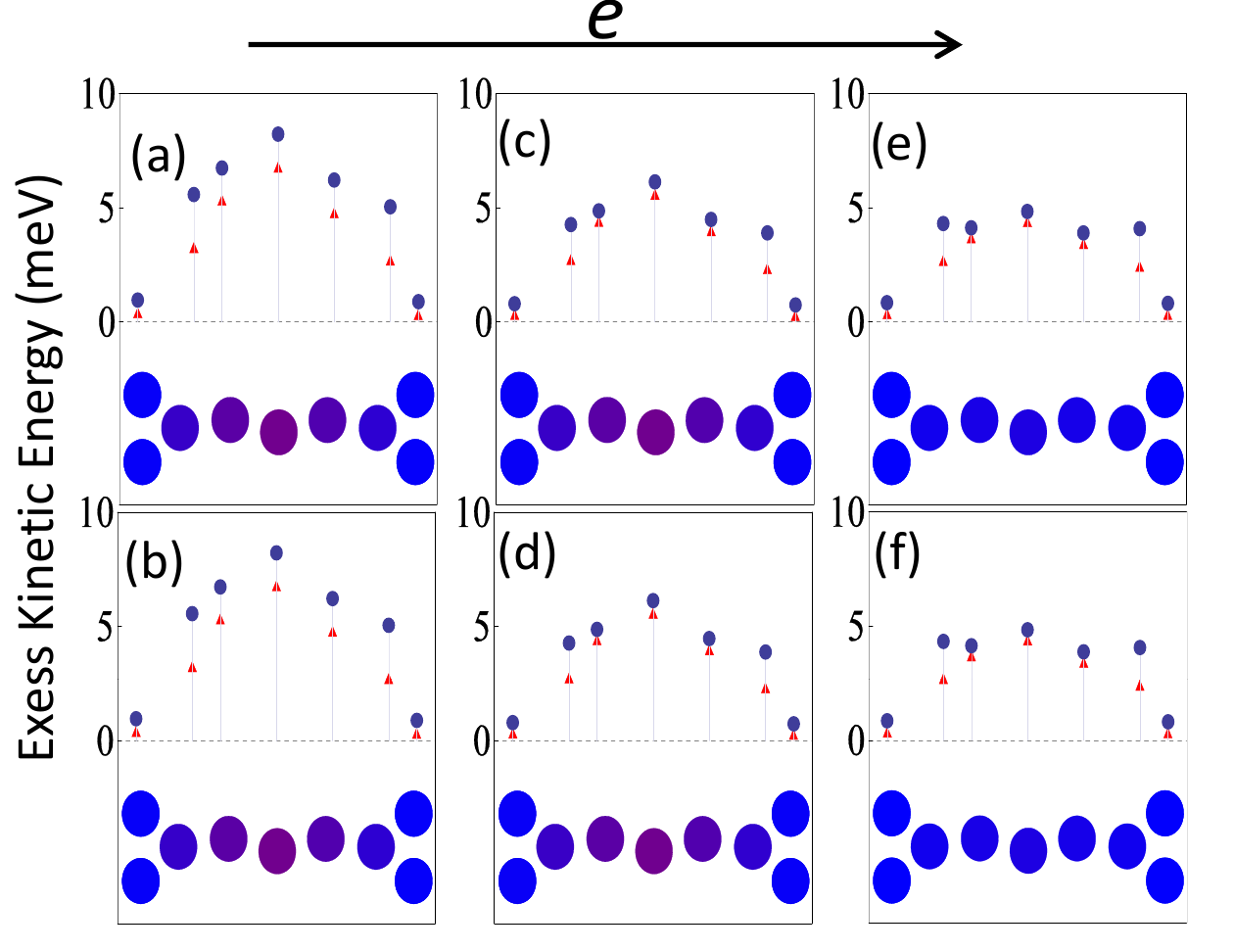}
\caption{Excess kinetic energy averaged over atoms in each column at $V=1.0$
V, $T=300$ K, similar to Fig.~2 in the main text. The top part (a), (c), (e)
shows results without the asymmetric current-induced forces, whiled the bottom
part (b), (d), (f) shows results that include only the BP and asymmetric
fluctuating force. This shows the contribution from the BP and fluctuating
force is negligible.}
\label{fig:excess2}
\end{figure}

\begin{figure}[!htbp]
\includegraphics[scale=0.5]{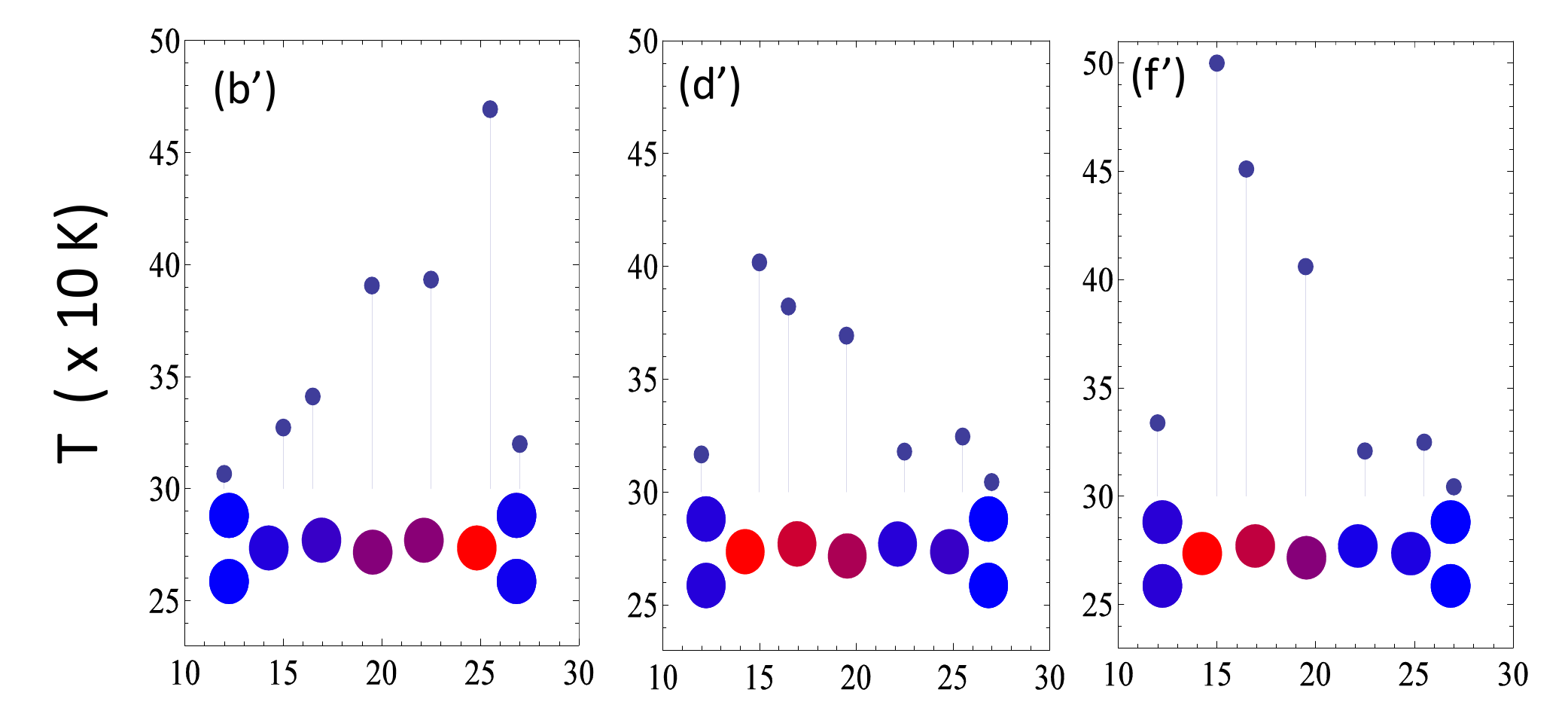}
\caption{Another way of characterizing heating in the chain is to use the
B\"uttiker probe (Refs. [40-44] of the main text) to `measure' the temperature
of each atom. (b'), (d') and (f') show the `measured' temperature of each atom using
this method when including all the forces. The overall heating profile agree with Fig. 2(b), (d), (f) in the main text.
}
\label{fig:excessa}
\end{figure}

\begin{figure}[!htbp]
\includegraphics[scale=0.6]{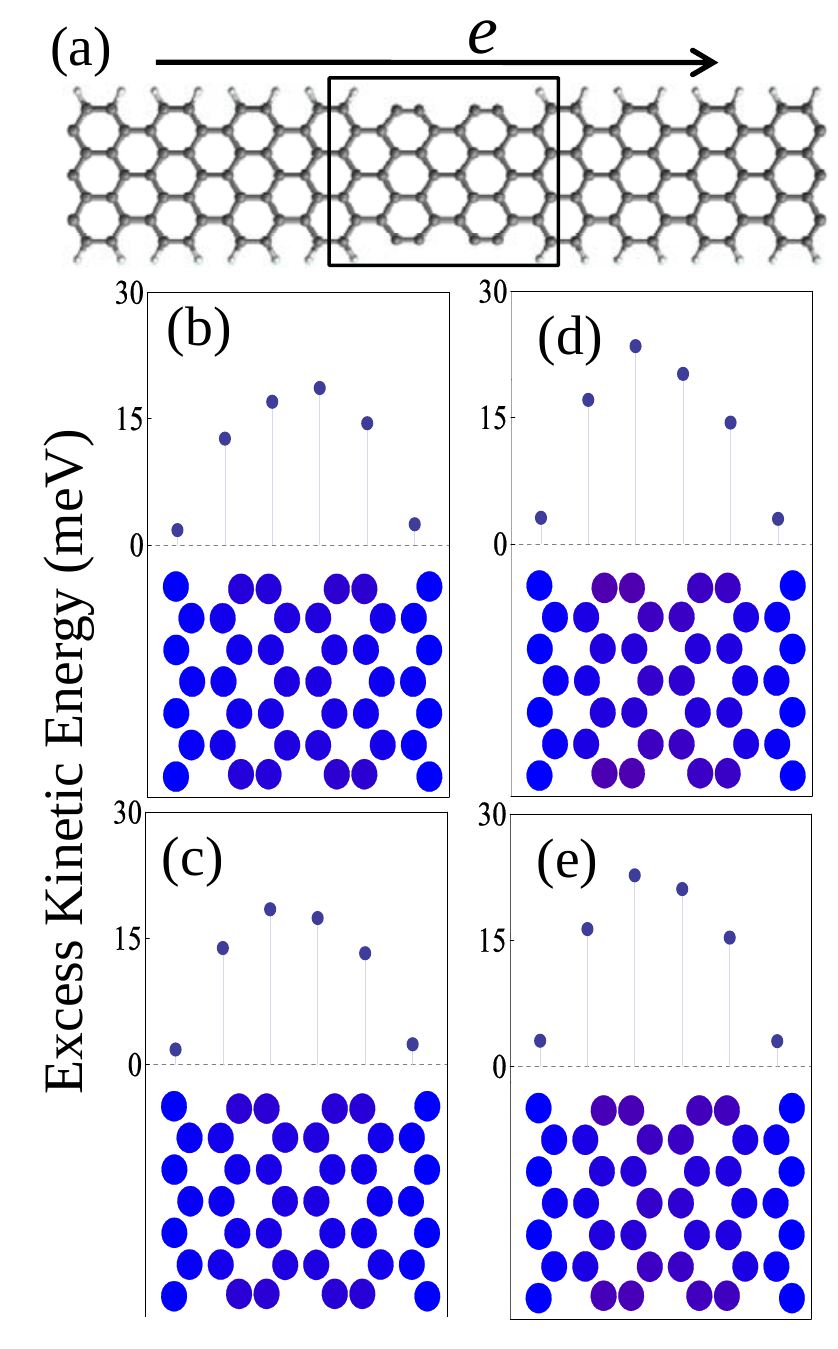}
\caption{Excess kinetic energy averaged over atoms in each column at $V=0.4$ V,
	$T=300$ K, similar to Fig.~3 in the main text. The top rows are results
	without the asymmetric current-induced forces, while the results in the bottom
	row include the BP and asymmetric fluctuating force. Again,
their contribution to the asymmetric heating is negligible.}
\label{fig:excess3}
\end{figure}

\begin{figure}[!htbp]
\includegraphics[scale=0.8]{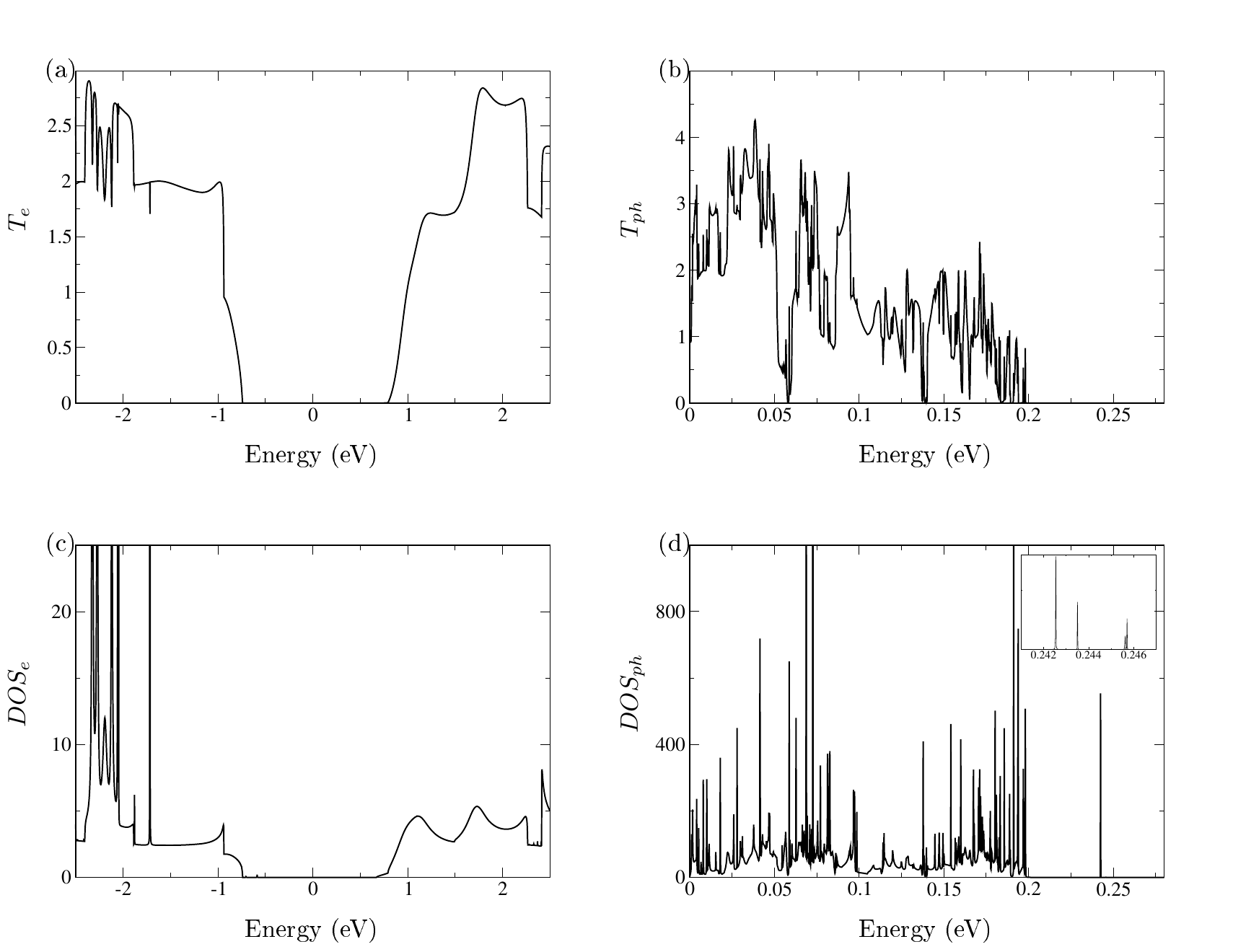}
\caption{Additional information on the Graphene nanoribbon calculation. (a) Electronic transmission ($T_e$). (b) Phononic transmission ($T_{ph}$). (c) Electronic density of states ($DOS_{e}$) projected to the device region. (d) Phononic density of states ($DOS_{ph}$) projected to the device region. In the SIESTA/TRANSIESTA DFT-calculation the following settings was used. Exchange-correlation functional: GGA-PBE. Basis-set: Single zeta polarized. Real space mesh cutoff: 400 Rydberg.  The structure was relaxed until the forces on the atoms in the device region was below 0.01 eV/Ang. }
\label{fig:TRANSandDOSaGNR}
\end{figure}

\newpage
\twocolumngrid

\end{document}